\documentclass[aps,prl,reprint,superscriptaddress]{revtex4-1}
\usepackage{amsfonts}
\usepackage{amssymb}
\usepackage{amsmath}
\usepackage{graphicx}
\usepackage{epsfig}
\usepackage{array}
\usepackage{braket}
\usepackage{multirow}
\usepackage[table,xcdraw]{xcolor}
\usepackage[colorlinks]{hyperref}

\begin{document}

\title{Cross-dimensional valley excitons from F\"{o}rster coupling in
arbitrarily twisted stacks of monolayer semiconductors}
\author{Ci Li}
\author{Wang Yao}
\email{wangyao@hku.hk}
\affiliation{Department of Physics, The University of Hong Kong, Hong Kong,
China}
\affiliation{HKU-UCAS Joint Institute of Theoretical and
Computational Physics at Hong Kong, China}
\begin{abstract}
In stacks of transition metal dichalcogenide monolayers with arbitrary
twisting angles, we explore a new class of bright excitons arising from the
pronounced F\"{o}rster coupling, whose dimensionality is tuned by its
in-plane momentum. The low energy sector at small momenta is
two-dimensional, featuring a Mexican Hat dispersion, while the high energy
sector at larger momenta becomes three-dimensional (3D) with sizable group
velocity both in-plane and out-of-plane. By choices of the spacer thickness,
interface exciton mode strongly localized at designated layers can emerge
out of the cross-dimensional bulk dispersion for a topological origin.
Step-edges in spacers can be exploited for engineering lateral interfaces to
enable interlayer communication of the topological interface exciton.
Combined with the polarization selection rule inherited from the monolayer
building block, these exotic exciton properties open up new opportunities
for multilayer design towards 3D integration of valley exciton
optoelectronics.
\end{abstract}
\maketitle


Atomically thin transition metal dichalcogenides (TMDs) have provided an
exciting platform for studying excitons in the two-dimensional (2D) limit.
Band-edge electrons and holes from the degenerate $\pm K$ valleys at the
Brillouin zone corners form tightly bound Wannier excitons with optical
selection rules for manipulating the valley pseudospin~\cite%
{Yao,Mak,Zen,Cao,Jon,Wan,Yao1,Wang,Hei,Hao}.
The small Bohr radius $\sim O(1)$ nm~\cite{Lou1,Rei,Cro}
further underlies a pronounced electron-hole (e-h) Coulomb exchange at
finite center-of-mass (COM) momentum that can transfer bright excitons
between these two valleys, with a coupling phase dependent on momentum
direction~\cite{Yao1,Wu,Yu,Song}. This realizes a sizable coupling between
exciton's valley pseudospin and COM degrees of freedom, splitting the
exciton dispersion into two branches having linearly polarized optical
dipole longitudinal ($L$) and transverse ($T$) to exciton momentum
respectively~\cite{Yao1,Yu,Lou,Mac,Thy,Abr}.
Notably, the $L$ branch is a massless one with group velocity proportional
to Coulomb strength, whose sensitive dependence on surrounding dielectric~%
\cite{Mal,Shan}
can lead to exotic properties of ground state excitons in a monolayer TMD on
patterned substrates~\cite{Yang1,Yang2}.


Reassembly of monolayer building blocks by van der Waals stacking further
provides versatile opportunities to explore exciton physics and
optoelectronic functionalities.
By its long-range nature, the e-h Coulomb exchange can also non-locally
transfer exciton or e-h pair, which is well known as the F\"{o}rster
coupling or F\"{o}rster energy transfer~\cite{For}. Efficient F\"{o}rster
energy transfer between excitons in a TMD monolayer and adjacent
nanostructures have been found~\cite{Tae}, promising sensing and imaging
applications~\cite{Jaz,Shao}. The F\"{o}rster coupling in principle allows
exciton to propagate out-of-plane in a stack of layers of homogeneous
excitonic resonance, even when charge hopping is completely quenched by
spacer layers or crystalline misalignment. This implies an intriguing yet
unexplored possibility to engineer the dimensionality of excitons separately
from that of charge carriers.

Here we discover a new class of bright excitons of cross-dimensionality,
introduced by the strong F\"{o}rster coupling in van der Waals stacks of
TMDs monolayers of quenched charge hopping. Exciton's in-plane COM momentum $%
k_{\parallel }$ serves as the parameter to tune its dimensionality, which
crosses from 2D at small $k_{\parallel }$ to 3D at large $k_{\parallel }$.
The low-energy sector features a series of 2D subbands well separated in
energy, where the energy minimum has a ring geometry. In the high-energy
sector, excitons acquire significant group velocity both in-plane and
out-of-plane. 
As the F\"{o}rster coupling concerns the momentum of exciton only, rather
than that of the electron or hole, the dispersion and polarization selection
rules of the cross-dimensional exciton are immune to the arbitrary rotation
of any layer in the stack. We further show that, by choices of the spacer
thickness, interface exciton mode strongly localized at designated layers
can emerge out of the cross-dimensional bulk dispersion, with a topological
origin that can be mapped to the Su-Schrieffer-Heeger (SSH) chain \cite%
{SSH,Pal}. Step-edges in spacers can be exploited for efficient interlayer
transfer of the topological interface exciton upon its lateral transport.
These findings point to a new avenue towards valley excitonic circuits of
multilayer design for high-level 3D integration.

\begin{figure*}[tbp]
\begin{center}
\includegraphics[width=0.75\textwidth]{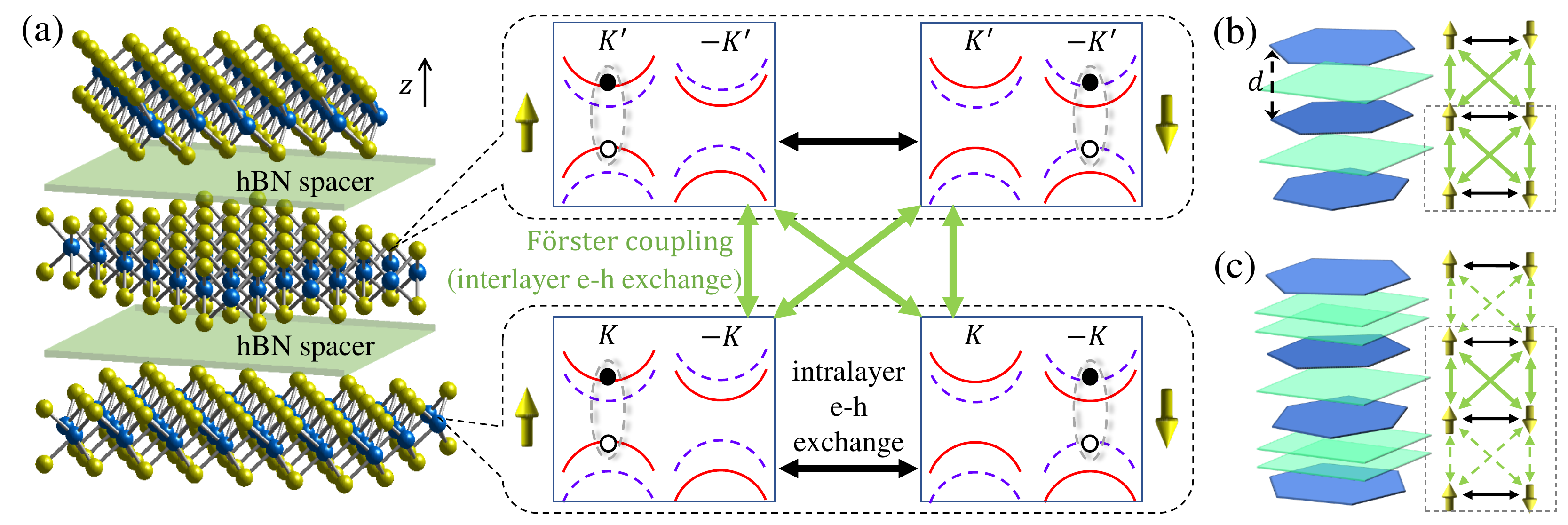}
\end{center}
\caption{(Color online)(a) Schematic of an arbitrarily twisted stack of
monolayer TMDs, where hBN spacers quench charge hopping. The two valley
configurations of bright excitons in each layer are coupled by the
intralayer electron-hole (e-h) Coulomb exchange (black arrows). Green arrows
denote the F\"{o}rster coupling, i.e., the interlayer e-h exchange, which
couples bright exciton states from different layers. Yellow arrows represent
the valley pseudospin. (b)-(c) Schematics of exemplary stacking
configurations, and the corresponding F\"{o}rster coupling between exciton
valley configurations from different TMD layers. As the F\"{o}rster coupling
conserves the exciton momentum in-plane, at each COM $k_{\parallel }$ the
exciton states form a quasi-one-dimensional (1D) chain. Only the nearest
neighbor F\"{o}rster coupling is displayed here, where the dashed arrow
denotes weaker coupling strength due to the thicker spacer.}
\label{fig1}
\end{figure*}

\begin{figure}[tbp]
\begin{center}
\includegraphics[width=0.38\textwidth]{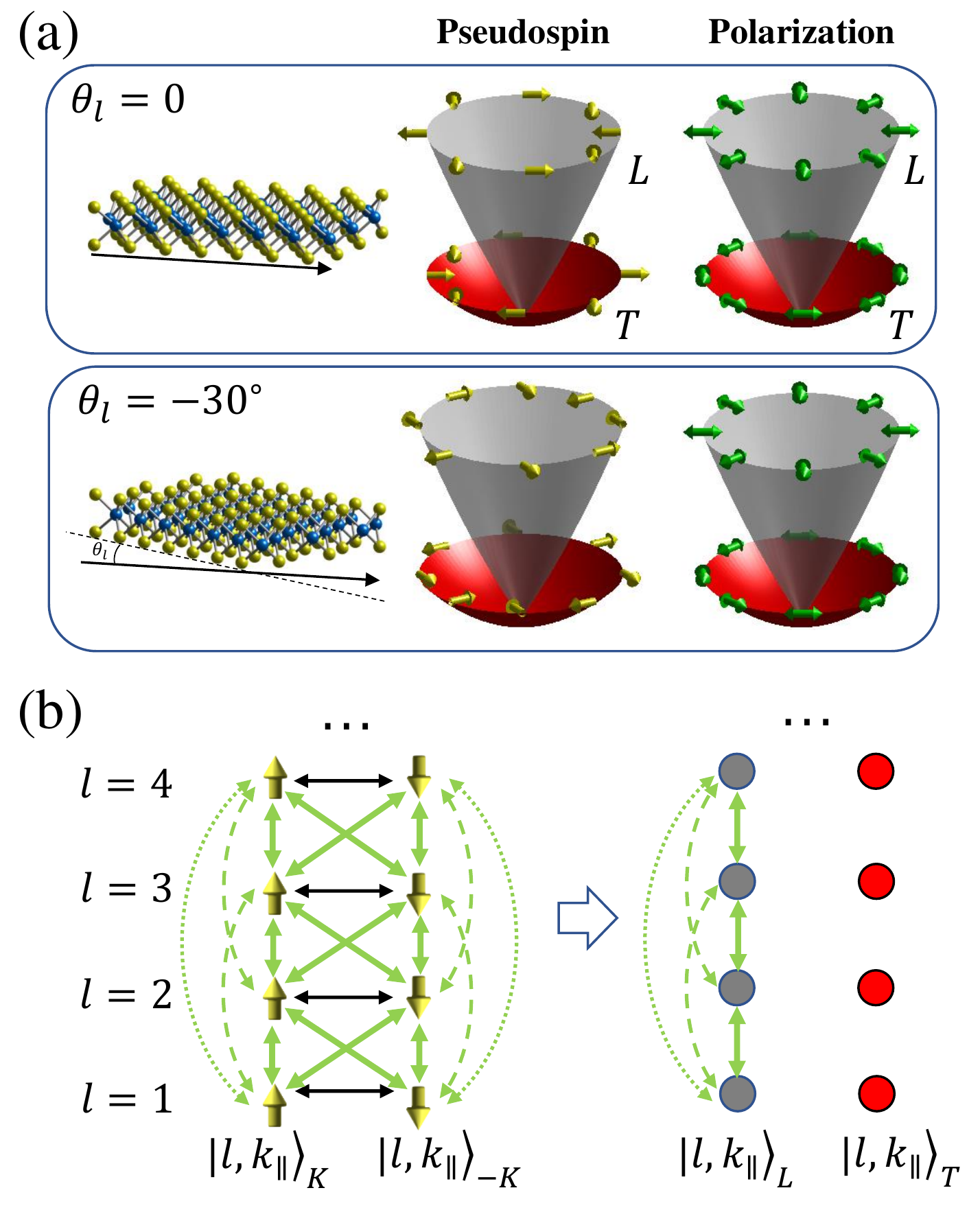}
\end{center}
\caption{(Color online)(a) $L$ and $T$ exciton branches split by the
electron-hole exchange in a monolayer TMD. The left column displays the valley
pseudospin texture (single-head arrow). In the gauge chosen (c.f. text), the
pseudospin texture rotates with the crystalline axis of the corresponding
layer. The right column displays the gauge-independent texture of the
polarization of the optical dipole (double-head arrow). (b) Exciton states
of a common COM momentum $\boldsymbol{k}_{\parallel }$ from different layers
are coupled nonlocally by the F\"{o}rster coupling, forming a 1D ladder in
the valley pseudospin basis. Transforming to the $L$ and $T$ basis, the $L$
states are coupled by the F\"{o}rster coupling into a chain, while all $T$
states are decoupled.}
\label{fig2}
\end{figure}

\begin{figure*}[tbp]
\begin{center}
\includegraphics[width=0.75\textwidth]{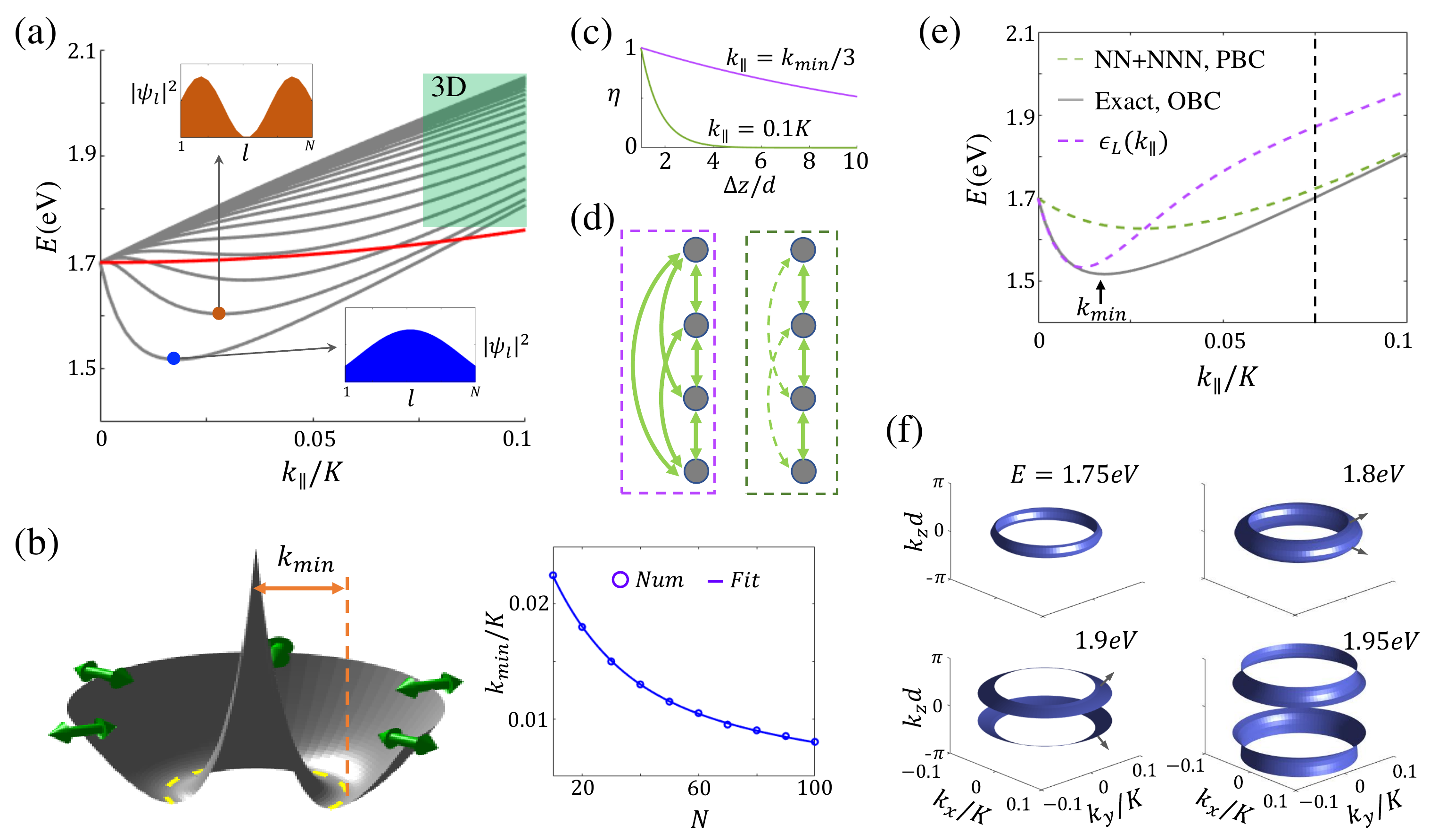}
\end{center}
\caption{(Color online)(a) Exciton dispersion for a stack of $N=20$ TMDs
layers with a uniform interlayer distance $d=1\,\mathrm{nm}$ (corresponding
to single layer hBN spacer). Grey (red) denotes the $L$ ($T$) branch
excitons (c.f. Fig. \protect\ref{fig2}(a)). F\"{o}rster coupling of $L$
excitons leads to a series of subbands well separated in the low energy
sector. Insets show layer distribution of exciton wavefunction in the two
lowest subbands. (b) Left: Mexican Hat dispersion of the lowest subband, $k_{%
\mathrm{min}}$ denoting the radius of the energy minima ring. Right: $k_{%
\mathrm{min}}$ as a function of the layer number $N$. (c) F\"{o}rster
coupling range indicated by the plot of $\protect\eta \equiv \exp
[k_{\parallel }(\Delta z-d)]$ as a function of out-of-plane distance $\Delta
z$, at two representative $k_{\parallel }$. (d) Chain of F\"{o}rster coupled
exciton states from different layers at a common $k_{\parallel }$. The
coupling is highly nonlocal at small $k_{\parallel }$ (left), compared to
that at large $k_{\parallel }$ (right). (e) The lowest exciton subband
(solid curve), in comparison with two approximations (dashed curves) for
small and large $k_{\parallel }$ regions respectively (Eqs.~(\protect\ref{ks}%
) and (\protect\ref{kl})). (f) Energy isosurfaces of the 3D exciton
dispersion in the high energy sector (green shaded in (a)), plotted using
Eq. (\protect\ref{hkl}) that keeps nearest-neighbor (NN) and next
nearest-neighbor (NNN) F\"{o}rster coupling. Arrows denote the direction of
exciton group velocity.}
\label{fig3}
\end{figure*}

\textit{Excitonic F\"{o}rster coupling in arbitrarily twisted multilayer} -
We consider stack of monolayers of the same TMD compound, all separated by
thin hBN spacers such that charge hopping is completely quenched (Fig.~\ref%
{fig1}). The Hamiltonian of valley excitons can be written as%
\begin{equation}
H=\frac{\hbar ^{2}k_{\parallel }^{2}}{2m_{X}}+\sum_{l}H_{\mathrm{intra}%
}^{l}+\sum_{l,l^{\prime }}H_{\mathrm{inter}}^{l,l^{\prime }}.  \label{H}
\end{equation}%
%
%
%
%
%
%
%
%
$\boldsymbol{k}_{\parallel }=\left( k_{\parallel }\cos \varphi ,k_{\parallel
}\sin \varphi \right) $ is the COM momentum in-plane. $H_{\mathrm{intra}}$
and $H_{\mathrm{inter}}$ denote respectively the intra- and inter-layer
electron-hole (e-h) Coulomb exchange.

$H_{\mathrm{intra}}$ consists of terms annihilating an e-h pair in one
valley while creating one in the same or opposite valley, with $\boldsymbol{k%
}_{\parallel }$ conserved. In the basis $\left\{ \left\vert l,\boldsymbol{k}%
_{\parallel }\right\rangle _{K},\left\vert l,\boldsymbol{k}_{\parallel
}\right\rangle _{-K}\right\} $ denoting exciton from valley $K$ or $-K$ of
layer $l$, it reads~\cite{Yao1,Yu,Lou,Mac,Thy,Abr},
\begin{equation*}
H_{\mathrm{intra}}^{l}=\left(
\begin{array}{cc}
J_{K,K} & J_{K,-K}^{l} \\
J_{-K,K}^{l} & J_{-K,-K}%
\end{array}%
\right) ,
\end{equation*}%
where
\begin{eqnarray}
J^l_{\pm K,\pm K} &\approx& \rho \left( 0\right) V\left( \boldsymbol{k}%
_{\parallel }\right) \left(e^{\mp i\theta_l} \boldsymbol{k}_{\parallel }\cdot \boldsymbol{d}%
_{cv,\pm K}\right) \notag \\
&&\times \left(e^{\mp i\theta_l} \boldsymbol{k}_{\parallel }\cdot \boldsymbol{d}%
_{cv,\pm K}\right)^{\ast }.
\end{eqnarray}
$\rho (0)\sim a_{B}^{-2}$ is the probability
for electron and hole to overlap in an exciton of Bohr radius $a_{B}$. $%
\boldsymbol{d}_{cv,\pm K}=d_{0}(\pm \hat{\mathbf{x}}-i\hat{\mathbf{y}})$ is
the optical transition dipole between conduction and valence band edges,
which is $\sigma +$ and $\sigma -$ polarized at $K$ and $-K$ respectively. $\theta _{l}$ is the twist angle of layer $l$ in the real space.
Taking an unscreened form for the Coulomb potential $V\left( \boldsymbol{k}%
_{\parallel }\right) $ then leads to $J^l_{K,K(-K,-K)}=J_{K,K(-K,-K)}=J\frac{k_{\parallel }%
}{K}$, where $K=4\pi /3a$, $a$ being TMD's lattice constant, and $J\sim 1$
\textrm{eV} can be extracted from first principle wavefunctions and exciton
spectrum~\cite{Yu,Lou,Lou1,Mac}. The off-diagonal term has the same
magnitude, but with a phase depending on the direction of $\boldsymbol{k}%
_{\parallel }$,
\begin{equation*}
J_{K,-K}^{l}=\left( J_{-K,K}^{l}\right) ^{\ast }=-J\frac{k_{\parallel }}{K}%
e^{-2i\left( \theta _{l}+\varphi \right) }.
\end{equation*}%
$\theta _{l}$ appears
as a global phase under the gauge choice that the valley pseudospin texture
in the eigen branches of $H_{\mathrm{intra}}^{l}$ rotates together with the
lattice (c.f. Fig.~\ref{fig2}(a), and Supplementary \cite{Supp}):
\begin{eqnarray}
\left\vert l,\boldsymbol{k}_{\parallel }\right\rangle _{L} &=&\frac{%
-e^{-i\left( \theta _{l}+2\varphi \right) }\left\vert l,\boldsymbol{k}%
_{\parallel }\right\rangle _{K}+e^{i\theta _{l}}\left\vert l,\boldsymbol{k}%
_{\parallel }\right\rangle _{-K}}{\sqrt{2}},  \notag \\
\left\vert l,\boldsymbol{k}_{\parallel }\right\rangle _{T} &=&\frac{%
e^{-i\theta _{l}}\left\vert l,\boldsymbol{k}_{\parallel }\right\rangle
_{K}+e^{i\left( 2\varphi +\theta _{l}\right) }\left\vert l,\boldsymbol{k}%
_{\parallel }\right\rangle _{-K}}{\sqrt{2}}.  \label{LTbasis}
\end{eqnarray}%
The optical dipoles of these eigenstates, nonetheless, are gauge invariant
quantities that remain longitudinal and transverse to $\boldsymbol{k}%
_{\parallel }$ in the two branches respectively~\cite{Yu,Bas}, regardless of
the twisting angle. This is guaranteed by the rotational symmetry of the
monolayer.

Likewise, the interlayer e-h exchange, i.e., F\"{o}rster coupling, consists
of $\boldsymbol{k}_{\parallel }$ conserving terms that annihilate an e-h
pair in layer $l$ while creating one in a different layer $l^{\prime }$. It
can be expressed in the same basis as,
\begin{eqnarray}
H_{\mathrm{inter}}^{l,l^{\prime }} &=&\left(
\begin{array}{cc}
J_{K,K}^{l,l^{\prime }} & J_{K,-K}^{l,l^{\prime }} \\
J_{-K,K}^{l,l^{\prime }} & J_{-K,-K}^{l,l^{\prime }}%
\end{array}%
\right) , \\
J_{K,K}^{l,l^{\prime }} &=&\left( J_{-K,-K}^{l,l^{\prime }}\right) ^{\ast
}\approx J\frac{k_{\parallel }}{K}e^{-k_{\parallel }\Delta z}e^{i\left(
\theta _{l^{\prime }}-\theta _{l}\right) },  \notag \\
J_{K,-K}^{l,l^{\prime }} &=&\left( J_{-K,K}^{l,l^{\prime }}\right) ^{\ast
}\approx -J\frac{k_{\parallel }}{K}e^{-k_{\parallel }\Delta z}e^{-i\left(
\theta _{l}+\theta _{l^{\prime }}\right) -2i\varphi }.  \notag
\end{eqnarray}%
The dependence on $\Delta z\equiv |\boldsymbol{z}_{l}-\boldsymbol{z}%
_{l^{\prime }}|$, i.e., vertical distance between layer $l$ and $l^{\prime }$%
, comes from the 3D form of Coulomb interaction~\cite{Kno} (see
Supplementary \cite{Supp}). From the $k_{\parallel }e^{-k_{\parallel }\Delta
z}$ dependence, we note that F\"{o}rster coupling can be neglected inside
the light cone where $k_{\parallel }$ is close to zero. At modestly small $%
k_{\parallel }$, a finite and highly non-local coupling developes, while at
large $k_{\parallel }$ it becomes short-ranged with large magnitude.

The total Hamiltonian in Eq.~(\ref{H}) now reads,
\begin{eqnarray}
H &=&\sum_{\lambda =L,T}\sum_{l}\varepsilon _{\lambda }\left\vert l,%
\boldsymbol{k}_{\parallel }\right\rangle _{\lambda }\left\langle l,%
\boldsymbol{k}_{\parallel }\right\vert _{\lambda }  \label{HL} \\
&&-2J\frac{k_{\parallel }}{K}\sum_{l,l^{\prime }}e^{-k_{\parallel }\Delta
z}\left( \left\vert l,\boldsymbol{k}_{\parallel }\right\rangle
_{L}\left\langle l^{\prime },\boldsymbol{k}_{\parallel }\right\vert
_{L}+H.c.\right),  \notag
\end{eqnarray}%
where $\varepsilon _{L}=\frac{\hbar ^{2}k_{\parallel }^{2}}{2m_{X}}+2J\frac{%
k_{\parallel }}{K}$ and $\varepsilon _{T}=\frac{\hbar ^{2}k_{\parallel }^{2}%
}{2m_{X}}$. Notably, in the basis of $L$ and $T$ polarized excitons (Eq.~(%
\ref{LTbasis})), $H$ has a gauge invariant form, independent of twist angles
$\{\theta _{l}\}$.

\begin{figure}[tbp]
\begin{center}
\includegraphics[width=0.45\textwidth]{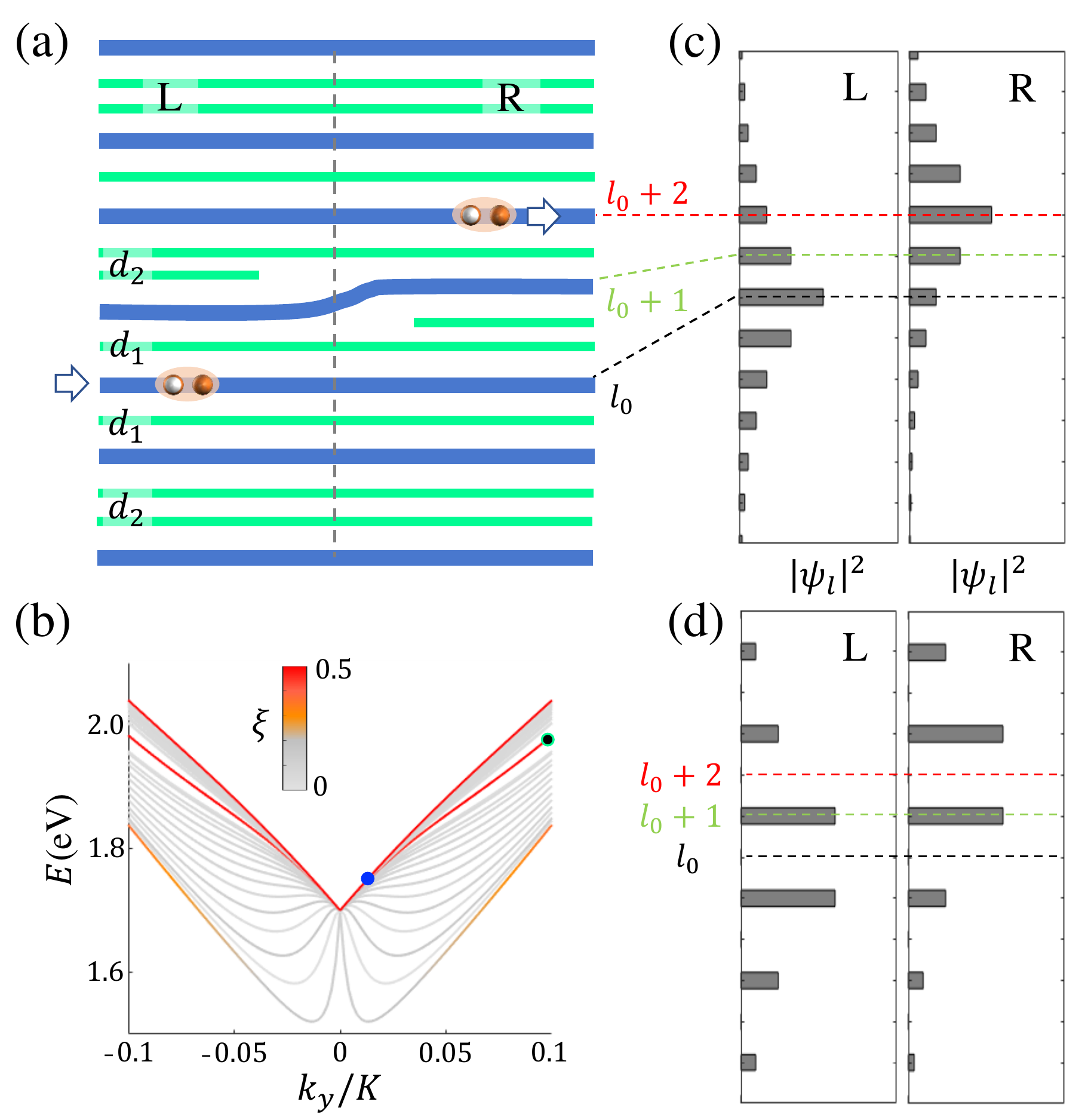}
\end{center}
\caption{(Color online)(a) Schematic of a TMD stack with an alternating
thickness of spacers, mimicking the Su-Schrieffer-Heeger (SSH) chain. Layer $%
l_{0}$ sandwiched by two spacers of $d_{1}$ thickness is a topological
interface in the SSH chain. A step-edge in the spacer layer defines a
lateral boundary, where the topological interface to its right is shifted
upward, becoming layer $l_{0}+2$. (b) The exciton dispersion of $L$ branch
in such a stack of $N=31$ TMDs, with a topological interface in the middle. $%
d_{1}=1\,\mathrm{nm}$ and $d_{2}=1.33\,\mathrm{nm}$, corresponding to one
and two hBN layers, respectively. Each subband is color-coded by $\protect%
\xi \equiv |\protect\psi _{l_{0}}|^{2}+|\protect\psi _{l_{0}-1}|^{2}+|%
\protect\psi _{l_{0}+1}|^{2}$, where $\protect\psi _{l}$ is the normalized
wavefunction on layer $l$. The red color highlights a topological interface
mode strongly localized on layer $l_{0}$. (c) The left (right) panel plots
the layer distribution of the interface mode to the left (right) side of the
step-edge, for the state marked by the blue dot on the dispersion in (b).
The center of the wavefunction is displaced by two TMD layers (c.f. (a)),
but still has a significant overlap to ensure efficient transmission across
the step-edge, realizing an interlayer communication. (d) A similar plot for
the state marked by the black dot in (b).}
\label{fig4}
\end{figure}

\textit{Cross dimensional valley exciton in the bulk} - At each $\boldsymbol{%
k}_{\parallel }$, the $L$ excitons from different layers are F\"{o}rster
coupled to form a chain (Fig.~\ref{fig2}(b)). The F\"{o}rster coupling range
is inversely proportional to $k_{\parallel }$, so distinct behaviors at
small and large momenta are expected. Fig.~\ref{fig3}(a) plots the
calculated exciton dispersions in a stack of $N=20$ TMDs layers, separated
by monolayer hBN spacer (c.f. Fig.~\ref{fig1}(b)). $T$ branch dispersion
(red curve) remains the same as that in monolayer, as they are not affected
by F\"{o}rster coupling (Fig.~\ref{fig2}(b)). In small $k_{\parallel }$
region, F\"{o}rster coupling splits the $L$ branch into a series of subbands
(grey curves) with energy separation of tens of meV. With such strong
quantization freezing the out-of-plane motion, exciton remains 2D, but the
dispersion is strongly renormalized. Remarkably, the lowest energy one
features a Mexican Hat dispersion, such that the exciton ground state
becomes highly degenerate (Fig.~\ref{fig3}(b)). The energy minima ring is
outside the light cone, where excitons can be scattered into the light cone
by long wavelength phonon ($k_{\mathrm{min}}\sim 10^{-2}K$), suggesting a
desirable balance between long lifetime and optical observability.

For $k_{\parallel }\leq k_{\min }$, we note that the F\"{o}rster coupling is
highly non-local (Fig.~\ref{fig3}(c)), and all layers in the $N=20$ stack
are strongly coupled to each other with roughly the same order of strength.
The coupling in the chain may be simplified as (Fig.~\ref{fig3}(d)),
\begin{equation*}
-J\frac{k_{\parallel }}{K}e^{-k_{\parallel }fd}\sum_{l,l^{\prime }}\left(
\left\vert l,\boldsymbol{k}_{\parallel }\right\rangle _{L}\left\langle
l^{\prime },\boldsymbol{k}_{\parallel }\right\vert _{L}+H.c.\right) ],
\end{equation*}%
which has analytical solution of the lowest subband as%
\begin{equation}
E_{0}\left( k_{\parallel }\right) =\varepsilon _{L}-2J\left( N-1\right)
\frac{k_{\parallel }}{K}e^{-fk_{\parallel }d}.  \label{ks}
\end{equation}%
With a factor $f=N/3.5$ to account for the weak distance dependence of the
long-range F\"{o}rster coupling, Eq.~(\ref{ks}) agrees well with the
numerical solution for $k_{\parallel }\leq k_{\mathrm{min}}$ (c.f. Fig. \ref%
{fig3}(e)).

For $k_{\parallel }\sim 0.1K$, the subbands in Fig. \ref{fig3}(a) get
densely packed so that motion in the out-of-plane becomes relevant.
Note that at such large $k_{\parallel }$, the F\"{o}rster coupling becomes
short ranged as shown in Fig. \ref{fig3}(c), which can be simplified by
keeping only the nearest and next-nearest neighbors (Fig.~\ref{fig3}(d))%
\begin{equation*}
-2J\frac{k_{\parallel }}{K}\sum_{s=1}^{2}e^{-sk_{\parallel }d}\left(
\left\vert l,\boldsymbol{k}_{\parallel }\right\rangle _{L}\left\langle l+s,%
\boldsymbol{k}_{\parallel }\right\vert _{L}+H.c.\right) ].
\end{equation*}%
It also has an analytical solution under periodic boundary condition (PBC),
\begin{eqnarray}
E\left( \boldsymbol{k}\right) &=&\varepsilon _{L}-4J\frac{k_{\parallel }}{K}%
(e^{-k_{\parallel }d}\cos k_{z}d+e^{-2k_{\parallel }d}\cos 2k_{z}d),  \notag
\label{hkl} \\
&&  \label{kl}
\end{eqnarray}%
In the region $k_{\parallel }\sim 0.1K$, it agrees well with the numerical
solution that keeps the full F\"{o}rster coupling under open boundary
condition (OBC) (Fig.~\ref{fig3}(e)).
In this high-energy sector, the dispersion becomes 3D, where exciton moves
with sizable group velocity that can reach $10^{5}\,\mathrm{m/s}$ both
in-plane and out-of-plane (Fig.~\ref{fig3}(f)).


\textit{Engineering topological interface exciton by choices of spacer
thickness} - The near-neighbor coupled chain structure in the large $%
k_{\parallel }$ limit suggests a design strategy to engineer topological
features using alternating thicknesses of spacers (Fig.~\ref{fig1}(c)). This
is reminiscent of the dimerized SSH chain, where nontrivial bulk winding
number leads to the emergence of topological modes at edges and interfaces~%
\cite{SSH,Pal}. The latter can be engineered by breaking the periodicity of
the spacer thickness pattern (see Fig.~\ref{fig4}(a)). Fig.~\ref{fig4}(b)
shows the calculated exciton dispersions of such stack with $N=31$ TMDs
monolayers, modes strongly localized at the interface (extended in
out-of-plane direction) are marked as red (grey) by color-coding.

Keeping only the nearest-neighbor F\"{o}rster coupling justifiable at
sufficiently large $k_{\parallel }$, for the stacking configuration in Fig.~%
\ref{fig4}(a) (see Supplementary \cite{Supp}), one can find one interface
mode in the bulk spectrum gap
\begin{equation*}
\left\vert \phi _{1}\right\rangle =\frac{1}{A}\sum_{s=1}\left( -\frac{v}{w}%
\right) ^{s-1}[\left\vert l_{\mathrm{dw}}-2s+1\right\rangle _{L}-\left\vert
l_{\mathrm{dw}}+2s-1\right\rangle _{L}]
\end{equation*}%
$l_{\mathrm{dw}}$ represents the interface layer (domain wall) as shown in
Fig.~\ref{fig4}(a). $A$ is the normalization factor, $w$ and $v$ denoting F%
\"{o}rster coupling strength in Eq. (\ref{HL}) at $\Delta z=d_{1}$ and $%
d_{2} $ respectively. There are also two interface modes at the top and
bottom edges of bulk spectrum,
\begin{eqnarray}
&&\left\vert \phi _{2}\right\rangle \approx \frac{1}{2}\left( \left\vert l_{%
\mathrm{dw}}+1\right\rangle _{L}+\left\vert l_{\mathrm{dw}}-1\right\rangle
_{L}\right) +\frac{1}{\sqrt{2}}\left\vert l_{\mathrm{dw}}\right\rangle _{L}
\notag \\
&&\left\vert \phi _{3}\right\rangle \approx \frac{1}{2}\left( \left\vert l_{%
\mathrm{dw}}+1\right\rangle _{L}+\left\vert l_{\mathrm{dw}}-1\right\rangle
_{L}\right) -\frac{1}{\sqrt{2}}\left\vert l_{\mathrm{dw}}\right\rangle _{L}
\notag
\end{eqnarray}%
From Fig.~\ref{fig4}(b), we can indeed identify the three modes at large $%
k_{\parallel }$. It is interesting to note that the two higher energy modes
remain strongly localized over the entire $k_{\parallel }$ region, even when
the above nearest-neighbor approximation breaks down at small $k_{\parallel
} $.

The possibility to engineer such topological interface mode at designated
layers further points to an interesting scenario by creating step-edges in
the spacer layers while keeping the TMDs layer intact (Fig.~\ref{fig4}(a)).
The wavefunction plotted in Fig.~\ref{fig4}(c,d) shows that the interface
modes at the two sides of the step-edge have their out-of-plane distribution
shifted by two TMDs layers. Their spatial overlap suggests a significant
lateral transmission probability through the step-edge, upon which the
exciton is also transferred across two layers. This can be exploited to
realize interlayer communication of valley excitons in a multilayer design
of valley excitonic circuits for 3D integration.


\textit{Acknowledgments} -C.L. would like to thank D. W. Zhai, B. Fu, H. Y. Zheng, and B. B for useful
discussions. This work is supported by the National Key R$\&$D Program of
China (2020YFA0309600), and Research Grant Council of Hong Kong SAR
(AoE/P-701/20,HKU SRFS2122-7S05). W.Y. also acknowledges support by Tencent
Foundation.


\begin{thebibliography}{99}
\bibitem{Yao} D. Xiao, G.-B. Liu, W. Feng, X. Xu, and W. Yao, Coupled spin
and valley physics in monolayers of MoS$_{2}$ and other group VI
dichalcogenides, Phys. Rev. Lett. \textbf{108}, 196802 (2012).

\bibitem{Mak} K. F. Mak, K. He, J. Shan, and T. F. Heinz, Control of valley
polarization in monolayer MoS$_{2}$ by optical helicity, Nat. Nanotech.
\textbf{7}, 494--498 (2012).

\bibitem{Zen} H. Zeng, J. Dai, W. Yao, D. Xiao, and X. Cui, Valley
polarization in MoS$_{2}$ monolayers by optical pumping. Nat. Nanotech.
\textbf{7}, 490--493 (2012).

\bibitem{Cao} Ting Cao, Gang Wang, Wenpeng Han, Huiqi Ye, Chuanrui Zhu,
Junren Shi, Qian Niu, Pingheng Tan, Enge Wang, Baoli Liu, and Ji Feng,
Valley-selective circular dichroism of monolayer molybdenum disulphide, Nat.
Commun. \textbf{3}, 887 (2012).

\bibitem{Jon} Aaron M. Jones, Hongyi Yu, Nirmal J. Ghimire, Sanfeng Wu,
Grant Aivazian, Jason S. Ross, Bo Zhao, Jiaqiang Yan, David G. Mandrus, Di
Xiao, Wang Yao, and Xiaodong Xu, Optical generation of excitonic valley
coherence in monolayer WSe$_{2}$, Nat. Nanotech. \textbf{8}, 634--638 (2013).

\bibitem{Wan} J. Kim, X. P. Hong, C. H. Jin, Su-Fei Shi, Chih-Yuan S. Chang,
Ming-Hui Chiu, Lain-Jong Li, and F. Wang, Ultrafast generation of
pseudo-magnetic field for valley excitons in WSe$_{2}$ monolayers, Science
\textbf{346}, 1205-1208 (2014).

\bibitem{Yao1} H. Yu, X. Cui, X. Xu, and W. Yao, Valley excitons in
two-dimensional semiconductors, Natl. Sci. Rev. \textbf{2}, 57--70 (2015).

\bibitem{Wang} G. Wang, X. Marie, B. L. Liu, T. Amand, C. Robert, F. Cadiz,
P. Renucci, and B. Urbaszek, Control of exciton valley coherence in
transition metal dichalcogenide monolayers, Phys. Rev. Lett. \textbf{117},
187401 (2016).

\bibitem{Hei} Z. Ye, D. Sun, and T. F. Heinz, Optical manipulation of valley
pseudospin, Nat. Phys. \textbf{13}, 26--29 (2016).

\bibitem{Hao} Kai Hao, Galan Moody, Fengcheng Wu, Chandriker Kavir Dass,
Lixiang Xu, Chang-Hsiao Chen, Liuyang Sun, Ming-Yang Li, Lain-Jong Li, Allan
H. MacDonald, and Xiaoqin Li, Direct measurement of exciton valley coherence
in monolayer WSe$_{2}$, Nat. Phys. \textbf{12}, 677--682 (2016).

\bibitem{Lou1} D. Y. Qiu, F. H. da Jornada, and S. G. Louie, Optical
spectrum of MoS2: many-body effects and diversity of exciton states, Phys.
Rev. Lett. \textbf{111}, 216805 (2013).


\bibitem{Rei} T. C. Berkelbach, M. S. Hybertsen, and D. R. Reichman, Theory
of neutral and charged excitons in monolayer transition metal
dichalcogenides. Phys. Rev. B \textbf{88}, 045318 (2013).

\bibitem{Cro} Andreas V. Stier, Kathleen M. McCreary, Berend T. Jonker,
Junichiro Kono, and Scott A. Crooker, Exciton diamagnetic shifts and valley
Zeeman effects in monolayer WS$_{2}$ and MoS$_{2}$ to 65\thinspace Tesla,
Nat. Commun. \textbf{7}, 10643 (2016).


\bibitem{Wu} T. Yu and M. W. Wu, Valley depolarization due to intervalley
and intravalley electron-hole exchange interactions in monolayer MoS$_{2}$,
Phys. Rev. B \textbf{89}, 205303 (2014).

\bibitem{Yu} H. Yu, G. Liu, P. Gong, X. Xu and W. Yao, Dirac cones and Dirac
saddle points of bright excitons in monolayer transition metal
dichalcogenides, Nat. Commun. \textbf{5}, 3876 (2014).

\bibitem{Song} Hanan Dery and Yang Song, Polarization analysis of excitons
in monolayer and bilayer transition-metal dichalcogenides, Phys. Rev. B
\textbf{92}, 125431 (2015).

\bibitem{Lou} Diana Y. Qiu, Ting Cao, and Steven G. Louie, Nonanalyticity,
Valley quantum phases, and lightlike exciton dispersion in monolayer
transition metal dichalcogenides: theory and first-principles calculations,
Phys. Rev. Lett. \textbf{115}, 176801 (2015).

\bibitem{Mac} F. Wu, F. Qu, and A. H. MacDonald, Exciton band structure of
monolayer MoS$_{2}$, Phys. Rev. B \textbf{91}, 075310 (2015).

\bibitem{Thy} T. Deilmann and K. S. Thygesen, Finite-momentum exciton
landscape in mono- and bilayer transition metal dichalcogenides, 2D Mater.
\textbf{6}, 035003 (2019).

\bibitem{Abr} D. Y. Qiu, G. Cohen, D. Novichkova, and S. Refaely-Abramson,
Signatures of Dimensionality and Symmetry in Exciton Band Structure:
Consequences for Exciton Dynamics and Transport, Nano Lett. \textbf{21},
7644 (2021).

\bibitem{Mal} A. Raja, L. Waldecker, J. Zipfel, Y. Cho, S. Brem, J. D.
Ziegler, M. Kulig, T. Taniguchi, K. Watanabe, E. Malic et al., Dielectric
disorder in two-dimensional materials, Nat. Nanotechnol. \textbf{14}, 832
(2019).

\bibitem{Shan} Y. Xu, S. Liu, D. A. Rhodes, K. Watanabe, T. Taniguchi, J.
Hone, V. Elser, K. F. Mak, and J. Shan, Correlated insulating states at
fractional fillings of moir\'{e}superlattices, Nature (London) \textbf{587},
214 (2020).

\bibitem{Yang1} Xu-Chen Yang, Hongyi Yu, and Wang Yao, Waveguiding valley
excitons in monolayer transition metal dichalcogenides by dielectric
interfaces in the substrate, Phys. Rev. B \textbf{104}, 245305 (2021).

\bibitem{Yang2} Xu-Chen Yang, Hongyi Yu, and Wang Yao, Chiral Excitonics in Monolayer Semiconductors on Patterned
Dielectrics, Phys. Rev. Lett. \textbf{128}, 217402 (2022).

\bibitem{For} Th. Forster, Energiewanderung und Fluoreszenz,
Naturwissenschaften \textbf{33}, 166--175 (1946).

\bibitem{Tae} Hyun Dong Ha, Dong Ju Han, Jong Seob Choi, Minsu Park, Tae
Seok Seo, Dual role of blue luminescent MoS$_{2}$ quantum dots in
fluorescence resonance energy transfer phenomenon, Small \textbf{10},
3858-62 (2014).

\bibitem{Jaz} A. Hichri, T. Amand, and S. Jaziri, Resonance energy transfer
from moir\'{e}-trapped excitons in MoSe$_{2}$/WSe$_{2}$ heterobilayers to
graphene: Dielectric environment effect, Phys. Rev. Materials \textbf{5},
114002 (2021).

\bibitem{Shao} Jie Zhou, Jiajie Chen, Yanqi Ge and Yonghong Shao,
Two-dimensional nanomaterials for F\"{o}rster resonance energy
transfer--based sensing applications, Nanophotonics 9, 7 (2020).

\bibitem{SSH} W. P. Su, J. R. Schrieffer, and A. J. Heeger, Solitons in
Polyacetylene, Phys. Rev. Lett. \textbf{42}, 1698 (1979).

\bibitem{Pal} J. K. Asb\'{o}th, L. Oroszl\'{a}ny, and A. P. P\'{a}lyi,
\textit{A Short Course on Topological Insulators} (Springer International
Publishing, Switzerland, 2016).

\bibitem{Kno} M. Selig, E. Malic, K. J. Ahn, N. Koch, and A. Knorr, Theory
of optically induced F\"{o}rster coupling in van der Waals coupled
heterostructures, Phys. Rev. B \textbf{99}, 035420 (2019).



\bibitem{Bas} L. C. Andreani and F. Bassani, Exchange interaction and
polariton effects in quantum-well excitons. Phys. Rev. B \textbf{41},
7536--7544 (1990).

\bibitem{Supp} See Supplementary for details regarding the electron-hole
Coulomb exchange, and analysis of SSH chain.



\end{thebibliography}
\end{document}